\documentclass[11pt]{article}
\usepackage[sfdefault]{carlito} % Use Carlito, similar to Calibri
\usepackage{geometry}
\usepackage{setspace}
\usepackage{titlesec}
\usepackage{hyperref}
\usepackage{graphicx}
\usepackage{amsmath}
\usepackage{bm}
\usepackage{wrapfig}
\usepackage{xcolor}
\usepackage{authblk}
\usepackage{soul}
\usepackage[font=small,labelfont=bf]{caption}

\usepackage{tabularx,booktabs} % for table
\usepackage[utf8]{inputenc}
\usepackage{amssymb}
\usepackage{cite}

\newcolumntype{Y}{>{\centering\arraybackslash}X}

\newcommand{\gi}[1]{\textbf{\textit{\textcolor{gray}{#1}}}}

\titleformat{\section}{\normalfont\fontsize{11}{13}\bfseries\sffamily}{\thesection}{1em}{}
\titleformat{\subsection}{\normalfont\fontsize{11}{13}\bfseries\sffamily}{\thesubsection}{1em}{}
\titleformat{\title}{\normalfont\fontsize{14}{16}\bfseries\sffamily}{}{0em}{}

\title{Solvers for Large-Scale Electronic Structure Theory: \\ ELPA and ELSI}
\author[1]{Petr Karpov}
\author[1]{Andreas Marek}
\author[1]{Tobias Melson}
\author[2]{Alexander P\"oppl}
\author[3]{Victor Wen-zhe Yu}
\author[4]{Ben Hourahine}
\author[5]{Alberto Garcia}
\author[6]{William Dawson}
\author[3,7]{Yi Yao}
\author[8]{William Huhn}
\author[9]{Jonathan Moussa}
\author[10]{Sam Hall}
\author[10]{Reinhard Maurer}
\author[3]{Uthpala Herath}
\author[7]{Konstantin Lion}
\author[7]{Sebastian Kokott}
\author[3,11]{Volker Blum}
\affil[1]{Max Planck Computing and Data Facility, Garching, Germany}
\affil[2]{Intel Corporation, Santa Clara, CA, USA}
\affil[3]{Thomas Lord Department of Mechanical Engineering and Materials Science, Duke University, Durham, NC 27708, USA}
\affil[4]{Department of Physics, SUPA, University of Strathclyde, John Anderson Building,107 Rottenrow, Glasgow G4 0NG, UK}
\affil[5]{Institut de Ciència de Materials de Barcelona, ICMAB-CSIC, Campus UAB, 08193 Bellaterra, Spain}
\affil[6]{RIKEN Center for Computational Science, Kobe 650-0047, Japan}
\affil[7] {Molecular Simulations from First Principles e.V., D-14195 Berlin, Germany}
\affil[8]{Intel Corporation, 500 Beaver Brook Road, Boxborough, MA, 01719, USA}
\affil[9]{Molecular Sciences Software Institute, Blacksburg, Virginia 24060, USA}
\affil[10]{University of Warwick, UK}
\affil[11]{Department of Chemistry, Duke University, Durham, NC 27708, USA}
\date{}

\begin{document}

\maketitle

\section*{Summary} % \\ \textit{\color{gray}(max 250 words)} } % 287 words

In this contribution, we give an overview of the ELPA library and ELSI interface, which are crucial elements for large-scale electronic structure calculations in FHI-aims. ELPA is a key solver library that provides efficient solutions for both standard and generalized eigenproblems, which are central to the Kohn-Sham formalism in density functional theory (DFT). It supports CPU and GPU architectures, with full support for NVIDIA and AMD GPUs, and ongoing development for Intel GPUs. Here we also report the results of recent optimizations, leading to significant improvements in GPU performance for the \textit{generalized} eigenproblem.

ELSI is an open-source software interface layer that creates a well-defined connection between ``user'' electronic structure codes and ``solver'' libraries for the Kohn-Sham problem, abstracting the step between Hamilton and overlap matrices (as input to ELSI and the respective solvers) and eigenvalues and eigenvectors or density matrix solutions (as output to be passed back to the ``user'' electronic structure code). In addition to ELPA, ELSI supports solvers including LAPACK and MAGMA, the PEXSI and NTPoly libraries (which bypass an explicit eigenvalue solution), and several others. 

ELSI, ELPA, and other solver libraries supported in ELSI are mature, well-tested software, ensuring efficient support for large-scale simulations on current HPC systems. Future plans for ELPA include further optimization of GPU routines, particularly integrating the GPU collective communication libraries (NVIDIA's NCCL and AMD's RCCL) into the solution tridiagonal of the matrix problem and backtransformation of the eigenvectors in standard eigenproblem solvers. Similarly expanding support for Intel GPUs through deeper integration with Intel's oneAPI libraries, specifically oneCCL, is planned. These advancements aim to ensure that ELPA, accessible either with its own Fortran/C/C++ APIs or via ELSI, remains at the forefront of large-scale computations, offering researchers powerful tools for addressing increasingly complex systems in electronic structure simulations. Future developments in ELSI will also continue to focus on GPU support and exascale readiness, continued support for a broad range of solvers, as well as extensions to solvers, e.g., for constrained density functional theory.

\section*{Current Status of the Implementation} % \\ \textit{\color{gray}(around 450 words)} } % 860 words

\gi{Introduction: From effective single-particle equations to generalized eigenproblem and density}

Using the Kohn-Sham (KS) \cite{kohn:1965} or generalized Kohn-Sham (gKS) \cite{Seidl1997} formalism, a full problem of $n_\mathrm{el}$ interacting electrons can be transformed to $n_\mathrm{el}$ or more auxiliary single-particle equations, $\hat{H}\psi_l = \varepsilon_l\psi_l$, e.g., for the scalar-relativistic or non-relativistic kinetic energy operator $\hat{t}$:
\begin{equation}
\left( \hat{t} + V_{\textrm{ext}}(\mathbf{r}) + \int \frac{n(\mathbf{r}')}{|\mathbf{r}-\mathbf{r'}|}d\mathbf{r}' + v_{\mathrm{xc}}(n(\mathbf{r}))\right) \psi_l (\mathbf{r}) =
\varepsilon_l \psi_l (\mathbf{r}) .
\label{full_schr_eq}
\end{equation}
The exchange-correlation potential $v_{\mathrm{xc}}$ is the only unknown term and can be treated using various density functional approximations \cite{Blum2024}. To solve the equations (\ref{full_schr_eq}) numerically, one has to choose a convenient set of $N$ 
% typically overlapping 
basis functions and approximate the KS state $\psi_l(\mathbf{r}) = \sum_{i=1}^N c_{li} \varphi_i (\mathbf{r})$. The set of equations (\ref{full_schr_eq}) then transforms to the generalized eigenvalue problem
\begin{equation}
H C = \varepsilon S C
\label{generalized_eigenproblem}
\end{equation}
where $H$ and $S$ are the Hamiltonian and the overlap matrices, respectively, with elements $H_{i,j} = \int \varphi_i (\mathbf{r})^* \hat{H} \varphi_j(\mathbf{r}) d\mathbf{r}$, $S_{i,j} = \int \varphi_i (\mathbf{r})^* \varphi_j(\mathbf{r}) d\mathbf{r}$; $C$ is the matrix of eigenvectors (stored as matrix columns); and $\varepsilon=\mathrm{diag}(\varepsilon_1,...,\varepsilon_N)$. Given $C$, one may obtain the total energy $E[n(\mathbf{r})]$ via the ground-state density
 \begin{equation}
 n(\mathbf{r}) = \sum_{l=1}^N f_l |\psi_l(\mathbf{r})|^2 ,
 \label{density}
 \end{equation}
where $f_l$ are the occupation numbers of the respective orbitals. 

\gi{Solving the generalized eigenproblem}

% math notation: $A X = \lambda B X$
% phys notation: $H C = E S C$
Since the basis functions $\{\varphi_i({\mathbf{r}})\}$ are not orthonormal, $S$ is not an identity matrix. In order to solve the generalized eigenproblem, $H C = \varepsilon S C$, for real symmetric/complex Hermitian matrices $H$, $S$, and positive-definite matrix $S$, one usually reduces it to the standard eigenproblem via the following steps:
\begin{enumerate}
\item
Cholesky decomposition of $S = U^T U$, where $U$ is an upper-triangular matrix.

\item
Inversion of the upper triangular matrix $U \rightarrow U^{-1}$.

\item
Calculation of two matrix products: $\tilde{H} = (U^{-1})^T H U^{-1}$.
\end{enumerate}
Here, $A^T$ corresponds to the transpose or conjugate-transpose of a real- or complex-valued matrix $A$, respectively. This reduces the \textit{generalized} eigenproblem to the \textit{standard} one, $\tilde{H} \tilde{C} = \varepsilon \tilde{C}$, which has the same set of eigenvalues. If the eigenvectors are needed, one has to perform the backtransformation, which requires another matrix-matrix multiplication: $C = U^{-1} \tilde{C}$.

A strict condition for the applicability of steps 1 to 3 is that the positive-definite matrix $S$ must not be numerically singular, i.e., the overlap matrix $S$ itself must not have eigenvalues that are zero; otherwise, Eq. (\ref{generalized_eigenproblem})) could have multiple different solutions $C$ that correspond to the same physical eigenfunctions $\{\psi_l(\mathbf{r})\}$. As a consequence, in practical computations, $S$ should not be ill-conditioned, i.e., the ratio between the largest and the smallest eigenvalue of $S$ should normally not exceed the numerical range (around $10^{12}$) accessible by double precision numbers. 

The standard FHI-aims basis sets are compact enough to lead to well-conditioned overlap matrices $S$ even for well-converged DFT calculations and steps 1 to 3 are therefore routinely used. Prior to any calculation, FHI-aims computes the eigenvalues $\sigma_i$ of the overlap matrix and alerts the user if a dense basis set with a high condition number of the overlap matrix is detected. In the latter case (high condition number of $S$), electronic structure calculations can then be carried out by transforming the eigenproblem (\ref{generalized_eigenproblem}) to the basis of eigenvectors $D_i$ of the overlap matrix, but omitting any $D_i$ for which $\sigma_i$ is smaller than a small positive number $\delta$ (typically $\delta \le 10^{-5}$). This procedure is standard in the community and is implemented in ELSI.

\gi{Circumventing the eigenvalue problem: Density-matrix based solutions}

In the limit of large systems, the computational effort for solving the eigenproblem Eq. (\ref{generalized_eigenproblem}) scales cubically with system size and becomes the computational bottleneck for Kohn-Sham DFT. Solving the eigenvalue problem (\ref{generalized_eigenproblem}) is one option to obtain $n(\mathbf{r})$. However, since the actual targets of DFT are $n(\mathbf{r})$ and $E[n(\mathbf{r})]$, the eigenvalues and eigenvectors are technically not required. Alternative strategies therefore target finding the single-particle density matrix $P_{i,j} = \sum_{l=1}^N f_l c^*_{li} c_{lj}$ without an explicit eigenvalue solution. The density can then be computed as 
  \begin{equation}
  n(\mathbf{r}) = \sum_{i,j} \int \varphi_i(\mathbf{r})^* P_{i,j} \varphi_j(\mathbf{r}) d\mathbf{r} . \label{density_dm}
  \end{equation}
Especially for sparse, localized basis sets, the computational effort for solvers that target the density matrix without an explicit eigenvalue/eigenvector solution can scale quadratically or linearly with the system size. For sufficiently large systems, such solvers can therefore outperform the cubic-scaling eigenvalue solution. The crossover point, i.e., the system size beyond which the eigenvector-free solution becomes more favorable, depends on the specifics of the system, level of theory, and computer hardware used. Thus, maintaining access to different solver types within a single electronic structure code is desirable.

\gi{ELSI}

\begin{figure}
    \centering
    \includegraphics[width=0.7\linewidth]{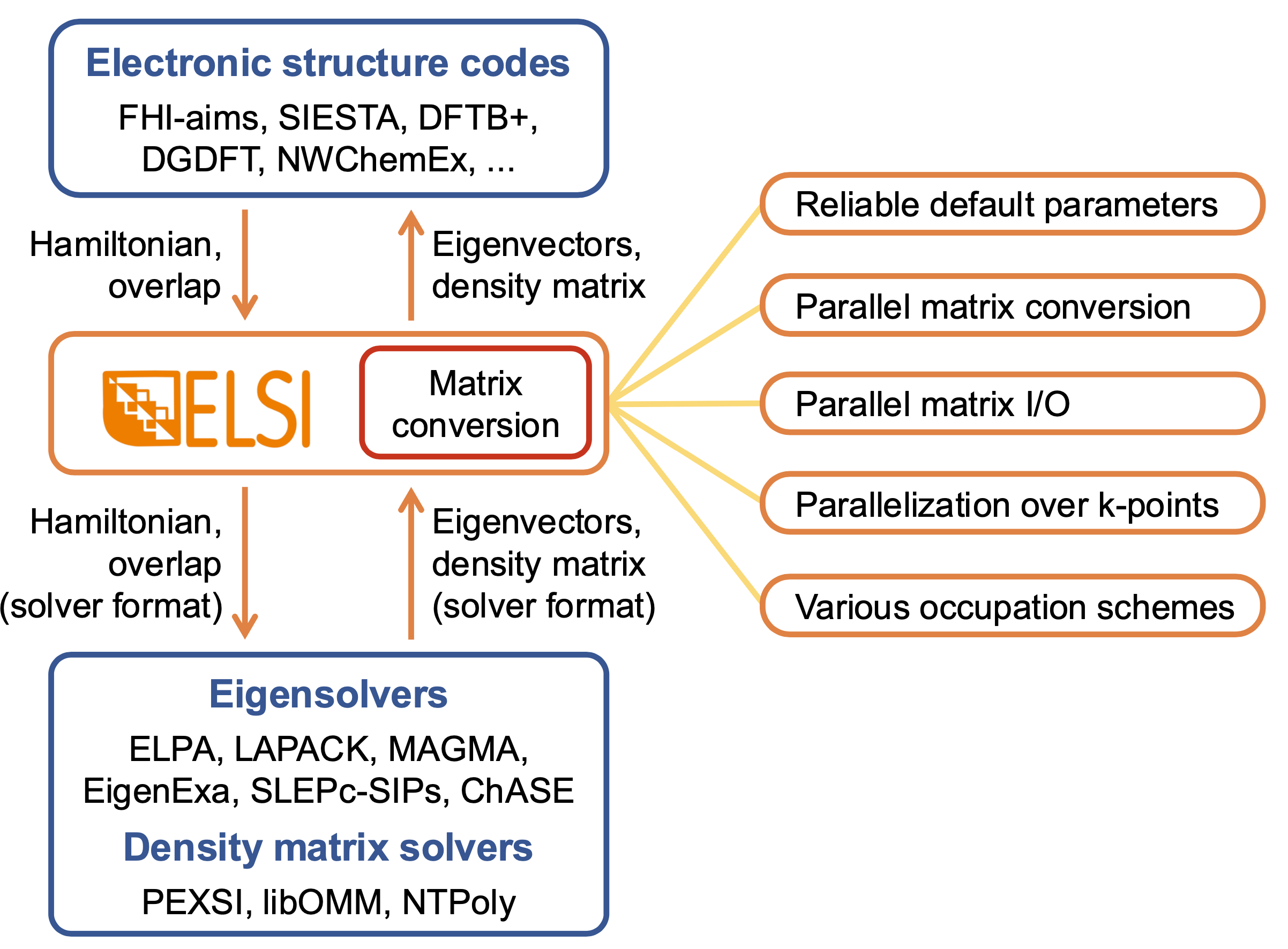}
    \caption{Tasks performed by the ELSI interface software, connecting different eigenvalue and density matrix solvers to electronic structure codes including FHI-aims, Siesta, DFTB+, NWChemEx and others. ELSI provides a uniform interface that is callable in Fortran, C, C++, and Python, and handles matrix format conversion between user codes and solver libraries. The solvers presented here include eigenvalue and density matrix solvers connected to ELSI at the time of writing. The ELSI interface provides functionality for tasks common to eigenvalue and density matrix operations, such as calculating occupation numbers or trivially parallel handling of independent eigenvalue problems for different $k$-points and spin channels.}
    \label{fig:ELSI}
\end{figure}

The ELSI interface software \cite{elsi_webpage,yu:2018,yu:2020} (Figure \ref{fig:ELSI}) provides a uniform code layer that handles eigenvalue problems or density matrix calculations, supporting several electronic structure codes (FHI-aims\cite{fhi-aims}, DFTB+\cite{dftb_plus}, Siesta\cite{siesta},NWChemEx\cite{NWChemEx}) and integrating ten different solvers.
The solvers have different APIs, making their direct integration into an electronic structure software a non-trivial task. 
Moreover, different electronic structure codes would all need to reimplement essentially the same connections to different solvers.
ELSI simplifies the integration and use of multiple solver libraries by providing a unified interface, allowing users to access various eigensolvers and density matrix solvers optimized for various problem sizes and types. The interface is designed around a derived type, the \texttt{elsi\_handle}, to pass data between the user code that calls ELSI, the ELSI interface layer itself, and the solver codes. This construct enables programs written in Fortran, C, C++ or Python to connect to the ELSI interface with equal ease. Additionally, conversions between different distributed-parallel matrix formats are handled efficiently in ELSI. 

ELSI supports cubic scaling eigensolvers like LAPACK \cite{netlib_lapack,anderson:1999} and ELPA \cite{elpa_webpage,marek:2014,marek:2024}, as well as reduced scaling methods such as PEXSI\cite{PEXSI-1,PEXSI-2} or the linear-scaling NTPoly\cite{NTPoly} density matrix solvers. Further supported solvers include the dense eigensolvers EigenExa \cite{EigenExa} and MAGMA\cite{MAGMA}, the iterative eigensolvers ChASE\cite{ChaSE1, ChaSE2}, SLEPc-SIPS\cite{SLEPc-SIPS}, and libOMM\cite{libOMM}, as well as BSEPACK \cite{BSEpack}.
ELSI returns either eigenvalues and eigenvectors or the density matrix and its energy-weighted counterpart. Since the density matrix includes the occupation numbers of the Kohn-Sham orbitals, ELSI also provides an implementation of numerically precise occupation numbers. For FHI-aims, ELSI is the designated source of standardized computed occupation numbers, $f_l$, including several non-Aufbau occupation approaches to perform occupation-constrained DFT calculations (e.g., the $\Delta$SCF method).

\gi{ELPA}

ELPA is an open-source massively parallel direct eigensolver library for real symmetric or complex Hermitian eigenvalue problems \cite{elpa_webpage,marek:2014,marek:2024}
that is supported by most major DFT software packages \cite{abinit,berkleyGW,cp2k,cpmd,dftb_plus,fhi-aims,gpaw,nwchem,openMX,quantumATK,quantum_espresso_1,quantum_espresso_2,siesta,vasp,wien2k}.
ELPA is capable of solving both \textit{standard} and \textit{generalized} eigenproblems: users are not required to perform the transformation from the \textit{generalized} problem manually, although interfaces for each individual transformation step are also provided.
Moreover, ELPA is ported to GPU accelerators (including NVIDIA, AMD, and Intel) \cite{{kus:2019_gpu,Yu:2021}}. 
%which can give a speedup factor of $\gtrsim 2-3$ on modern hardware, in comparison to the best setting for CPU-only nodes.
To the best of our knowledge, ELPA has recently set a new record \cite{wlazlowski:2024} of a $3.2 \cdot 10^6 \times 3.2 \cdot 10^6$ dense matrix diagonalization for all eigenvalues and eigenvectors of the standard real-valued eigenproblem on the LUMI supercomputer, which is equipped with AMD MI250X GPUs.

\section*{Usability and Tutorials} % \\ \textit{\color{gray}(around 450 words)}} % 670 words

\gi{Solver selection in FHI-aims}

The numerical atomic orbital (NAO) basis sets in FHI-aims \cite{fhi-aims} are rather compact, i.e., the number of basis functions $N$ required to represent the single-particle states with high precision is typically not larger than a single-digit multiple of $n_\mathrm{el}$. Thus, since the number of basis functions $N$ determines the dimension of $H$ and $S$, typically a large fraction of their eigenpairs is needed.
For this scenario and for small and mid-sized systems, direct eigensolvers, such as LAPACK, ScaLAPACK \cite{netlib_scalapack,blackford:1997} or ELPA are the solvers of choice. 

One advantage of ELSI is its ability to enable direct, comparative benchmarks of different solvers within a single software framework, allowing one to make a choice that is specific to system and hardware characteristics of a particular computational campaign.
Beyond the default serial (LAPACK) and parallel (ELPA) solvers available in FHI-aims, a range of additional solvers can be accessed through ELSI. At the time of writing, these include PEXSI, NTPoly, EigenExa, SLEPc-SIPS, libOMM, and MAGMA.

Past results for the NAO basis sets of FHI-aims showed that the crossover point beyond which different density-matrix-based solvers outperform ELPA lies beyond several hundred to several thousand atoms \cite{yu:2020}. On parallel and/or GPU hardware, performing a direct eigenvalue solution using ELPA is therefore preferable for the majority of current production calculations in FHI-aims. For larger systems and for semilocal DFT, the solution effort for the pole expansion and selected inversion (PEXSI)\cite{PEXSI-1,PEXSI-2} method scales at most quadratically with system size for three-dimensional systems, and even subquadratically for two- or one-dimensional geometries. Benchmarks in Ref. \cite{yu:2020} showed that, for 1D systems, PEXSI outperformed a direct diagonalization for as few as 400 atoms. For 2D systems, the crossover was found around 1,000 atoms. Alternatively, for non-metallic systems, density matrix purification allows one to directly solve for the density matrix with a solution effort that grows linearly with the system size, independent of the density functional, as implemented, e.g., in the NTPoly\cite{NTPoly} library for sparse matrix function calculations. 

\gi{ELPA - Usability}

Within FHI-aims, ELSI and ELPA (as well as LAPACK calls for serial linear algebra steps) are automatically employed with parallelization strategies that leverage trivial parallelism (e.g., for separate $k$-points or spin channels) whenever possible. Since the performance of solvers other than ELPA depends on system, problem and hardware characteristics, their selection is a user choice by setting specific keywords.

One key consideration pertains to high-performance hardware beyond CPUs. Specifically, utilizing GPUs is essential for leveraging the computational power of modern supercomputers.
This not only accelerates matrix diagonalization but also allows for the handling of larger matrices,
enabling the investigation of physical systems with more atoms.

Much effort has been spent porting ELPA to GPUs. Nowadays ELPA fully supports both NVIDIA and AMD GPUs, including support for their respective GPU collective communication libraries, NCCL \cite{nccl} and RCCL \cite{rccl}, respectively, for the most computationally intensive parts of the algorithm.
Partial support is available for Intel GPUs using SYCL \cite{intelsycl} and oneMKL \cite{intelonemkl}, though this is currently limited to solving the standard eigenproblem.

For the core solver of the \textit{standard} eigenproblem, ELPA provides implementations of the conventional one-stage diagonalization method  (``ELPA1'') and the two-stage diagonalization (``ELPA2'') \cite{lang:1993, lang:1994, auckenthaler:2011}.
On CPUs, as a rule of thumb, the ELPA2 solver is preferable and it is also typically up to 1.5-2 times faster than MKL's ScaLAPACK \cite{kus:2019}.
However, on GPUs, the ELPA1 solver is typically faster, providing a speedup of three times over the best setting on the CPU-only nodes if the local matrices per GPU are not extremely small \cite{kus:2019_gpu}. If only a subset of all eigenvectors is computed (as is often the case in DFT), then ELPA2-GPU becomes more competitive compared to ELPA1-GPU as the number of targeted eigenvectors decreases.

For the \textit{generalized} eigenproblem, until recently ELPA-GPU faced a significant bottleneck in the matrix-matrix multiplication step.
ELPA relied on ScaLAPACK or its own implementations of the SUMMA \cite{de_geijn:1997,chtchelkanova:1997} and Cannon's \cite{cannon:1969} algorithms for parallel matrix multiplication, none of which were fully GPU-ported until the 2024.05 release, leaving a substantial part of the computation to CPUs.
Table~\ref{table:elpa-generalized} compares the ELPA 2023.11 and 2024.05 releases, with the latter including a full GPU port for the matrix-matrix multiplication (`Multiply'). This is especially important for `NCCL' setup (1 MPI task per GPU), where the total `Multiply' time for forward and backward transformation is reduced from 559 to 25.6 seconds. 

GPU porting of the matrix-matrix multiplication brings the performance of the ELPA-GPU \textit{generalized} eigensolver in line with that of the core \textit{standard} eigenvalue solver, now achieving a similar speedup of approximately three to four times over the best-performing ELPA-CPU configuration. It is important to note that the CPU side of the comparison (last row of Table~\ref{table:elpa-generalized}) involves all CPU cores on a given computational node. For instance, as shown in Table~\ref{table:elpa-generalized}, we compare a GPU setup with 4 GPUs and tuned from 1 to 18 CPU cores per GPU (using ELPA1) against the CPU setup with all 72 CPU cores (using ELPA2). At the moment, FHI-aims is best tested with NVIDIA MPS \cite{mps}, as it allows multiple MPI processes per GPU to be used efficiently, which can be particularly beneficial for steps like `Solve'. NVIDIA MPS using all CPU cores of a node seems still to provide a very good speedup compared to the CPU-only ELPA (cf.~Table~\ref{table:elpa-generalized}). In cases where ELPA dominates the FHI-aims runtime, it is worthwhile to tune for the best MPI process per GPU ratio. For the future development of ELPA, as more of its parts being ported to GPU over time, we expect that the `NCCL' code path will become more performant than the `MPS' one.
This is because NCCL enables direct data transfer between GPU devices, eliminating the need for costly GPU-CPU memory copies required for MPI operations.

Overall, we recommend using ELPA-GPU when possible: it can provide up to a 4x speedup for the complete solution of standard and generalized eigenproblems, and even up to a 10x speedup for individual solution steps.
Despite this drastic improvement provided by the ELPA 2024.05 release, the `Multiply' step still dominates the time for forward and back transformations from \textit{generalized} to \textit{standard} eigenproblem, motivating its further optimization.

ELSI supports ELPA-GPU (NVIDIA only) directly in ELPA's 2020 release, and as an externally compiled library for later versions and other GPU types. Since newer ELPA versions offer significant performance enhancements on GPUs, it is recommended to use ELSI with externally linked ELPA.

\begin{table}[t]
\centering
\begin{tabularx}{\textwidth}{c || *{7}{Y} || Y}
\toprule
& \multicolumn{3}{c}{Forward transformation} & \multicolumn{3}{c}{Standard EVP} & Backtr. &  \\
\cmidrule(lr){2-4} \cmidrule(lr){5-7} \cmidrule(lr){8-8} %\cmidrule(lr){9-9}
ELPA version               & Choles.  & Invert  & Multiply        & Tridiag.  & Solve  & Back     & Mult.          &  Total \\
\midrule
2023.11, MPS, 18 MPI per GPU  & $6.5$ & $5.7$   & $\mathbf{34.7}$ & $55.3$    & $11.7$ & $18.0$   & $\mathbf{24.5}$& $156.4$ \\
2024.05, MPS, 18 MPI per GPU  & $6.1$ & $4.9$   & $21.4$          & $54.5$    & $12.1$ & $17.6$   & $12.1$         & $128.9$ \\
2024.05, MPS, 4  MPI per GPU  & $5.9$ & $5.1$   & $8.0$           & $51.3$    & $11.3$ & $12.6$   & $3.3 $         & $97.8$ \\
%/home/pekarp/raven-i/git/benchmarks/elpa_2024_05/build_openmpi_4_1_cuda_11_4_nvtx_ucx_1_15/runs_generalized_evp_cannons/plain/run_plain,gpu=4_N=40960,optimization_1_mps,4_mpi_per_gpu_15020635/
\midrule
2023.11, NCCL, 1 MPI per GPU  & $2.8$ & $2.1$   & $\mathbf{294}$  & $49.7$    & $40.4$ & $15.3$   & $\mathbf{265}$ & $669$ \\
2024.05, NCCL, 1 MPI per GPU  & $2.3$ & $1.8$   & $18.8$          & $49.5$    & $40.1$ & $14.6$   & $6.8 $         & $134$ \\
%/home/pekarp/raven-i/git/benchmarks/elpa_2024_05/build_openmpi_4_1_cuda_11_4_nvtx_nccl_ucx_1_15/runs_generalized_evp_cannons/plain/run_plain,gpu=4_N=40960,optimization_0_11900036/
\midrule
2024.05, CPU & $27.3$ & $23.4$   & $72.7$         & $54.4$    & $42.9$ & $141$    & $30.3$         & $393$ \\
\bottomrule
\end{tabularx}
\caption{Comparison of the wall-clock times (in seconds) of the ELPA 2023.11 vs 2024.05 releases in different GPU and CPU setups for the individual steps of the \textit{generalized} eigenproblem.
The matrix size is $40960\times40960$, and the diagonalization is performed to obtain all eigenvalues and eigenvectors on a single node on MPCDF's system \textit{Raven} \cite{raven}.
For the GPU calculations the ELPA1 solver with four NVIDIA A100 (40GB, SXM) GPUs was used. The `MPS' configuration uses NVIDIA's Multi-Process Service \cite{mps} and several MPI processes per GPU. 
For the ELPA 2023.11 release, 18 MPI processes per GPU (full Raven node) show the best performance, while for ELPA 2024.05, the best performance is achieved with 4 MPI processes per GPU.
%tuned for best performance (18 and 4 MPI processes per GPU for 2023.11 and 2024.05 ELPA releases respectively).  % and NB=1024
The `NCCL' configuration uses the NVIDIA Collective Communication Library \cite{nccl}, which is constrained to 1 MPI process per GPU.
For the CPU calculations the ELPA2 solver with 72 MPI processes on two 36-core Intel Xeon Platinum 8360Y processors was used. % and NB=32
%\rr{[Values for the upcoming ELPA 2024.11 (pre-) release are to be added in Nov-Dec 2024]}
}
\label{table:elpa-generalized}
\end{table}

\gi{Tutorials}

Within the FHI-aims ecosystem, ELPA and ELSI are automatically used in standard tutorials, however, ELSI itself comes with its own extensive manual in its repository. Within the standard build process of FHI-aims, the 2020.05.001 version of ELPA is included as a default build step. A newer version of ELPA can be included first by compiling an external version of the ELPA library and then linking to this version in a subsequent build of FHI-aims; this process is described as a dedicated tutorial accessible at \url{https://fhi-aims-club.gitlab.io/tutorials/tutorials-overview/}.

Since the 2024.05 release, ELPA provides a comprehensive and self-contained \textit{ELPA Manual: User's Guide and Best Practices} \cite{marek:2024}. This manual offers ELPA's quick-start guide and code examples, as well as detailed instructions on installation, how to use ELPA in applications, and troubleshooting.
It is an essential resource for both new and experienced users, ensuring optimal performance and integration of ELPA in applications.
Additionally, we provide a GitHub repository containing teaching materials from recent tutorials on ScaLAPACK and ELPA \cite{karpov:2023}. The repository includes presentation slides and various code examples that demonstrate the practical use of ELPA, covering both the CPU and GPU versions of the library.

\section*{Future Plans and Challenges} % \\ \textit{\color{gray}(around 350 words)}} % 267 words

Several key optimizations are planned for ELPA to further improve its performance on modern HPC systems.
The primary focus is on enhancing the GPU solvers for the standard eigenproblem, particularly focusing on the ELPA2 tridiagonalization, the solve step of the tridiagonal matrix, and the backtransformation steps, which are not yet utilizing the GPU collective-communication libraries. This should allow the code to keep all the data on the GPU memory, avoiding the costly data transfers between the CPU and GPU. Overall, we expect that with more and more parts of ELPA being ported to GPU, using NCCL/RCCL/oneCCL collective communication libraries allowing direct GPU-GPU communication will become more advantageous; in particular, we expect that ELPA's `NCCL' codebranch will outperform the `MPS' one in the near future. However, the use of NCCL and alike will require a reorganization of the management of the GPU use in FHI-aims: it is necessary to set up the dense matrices directly in the GPU memory corresponding to a given MPI task, instead of first distributing them in parallel across all CPUs.

Additionally, work is underway to optimize ELPA's parallel matrix-matrix multiplication routine, which is used for the transformation of the generalized to the standard eigenproblem. Although the GPU implementation has already reduced the bottleneck as we reported in this contribution, the matrix multiplication remains the most time-consuming part of the process.

Expanding support for Intel GPUs is another major goal. Currently, ELPA's functionality on Intel GPU hardware is limited to solving the standard eigenproblem. We plan to utilize oneCCL and further oneMKL operations to achieve feature parity with the other GPU implementations in ELPA.

Generally, ELPA aims to support and be optimized for all modern and emerging hardware architectures, including Accelerated Processing Units (APUs) \cite{apu} and Vector Processing Units (VPUs) \cite{vpu}. Notably, there is an ongoing effort to port ELPA kernels to the RISC-V architecture \cite{risc-V-elpa}.

ELSI will need similar architecture-specific optimizations. Additionally, support for a broader ecosystem of new and emerging solvers is a constant target, with recent, community-supported additions of the Chebyshev eigensolver ChaSE \cite{ChaSE1,ChaSE2} and a new, distributed linear algebra eigensolver DLA-Future \cite{DLA-Future,Solca:2024}. For example, demonstration calculations of extremely large DFT calculations were performed on Google's proprietary Tensor Processing Units (TPUs) \cite{Pederson:2022}, based on an adaptation of the FHI-aims/ELSI software stack. This adaptation also made use of an efficient combination of single- and double-precision solutions during the self-consistent field cycle. Similar use of faster single-precision solutions in ELPA is possible using a reduced, frozen-core eigenvalue solution via ELSI and FHI-aims, offering further optimization potential in the future \cite{Yu2021fc}.

\section*{Acknowledgements}

We are grateful to Marc Torrent for sharing his preliminary ELPA benchmarks with ABINIT and for fruitful discussions, which have triggered the work on optimizing the generalized eigenproblem solver in ELPA-GPU.
We gratefully acknowledge the support provided by Markus Hrywniak from NVIDIA.
We also thank Markus Rampp for the useful comments on the initial draft of the manuscript.
This project was supported by NOMAD Center of Excellence
(European Union's Horizon 2020 research and innovation program, Grant Agreement No. 951786) and the embedded CSE programme of the ARCHER2 UK National Supercomputing Service (http://www.archer2.ac.uk).
We acknowledge CSC for awarding us access to LUMI supercomputer (Kajaani, Finland).
ELSI was supported by the National Science Foundation (NSF), USA under Award No. 1450280.
We also thank all ELSI authors who contributed to ELSI in the past and/or in the context of other projects, specifically: Bj{\"o}rn Lange, Ville Havu, Jianfeng Lu, Lin Lin, Fabiano Corsetti, Alvaro Vazquez-Mayagoitia, Weile Jia, Raul Laasner, Yingzhou Li, David B. Williams-Young, Xinzhe Wang, Edoardo di Napoli, and Rocco Meli.

% Explicit bibliography to make it easier to format in the future

\end{document}